\documentstyle[aps,multicol,epsf]{revtex}
\narrowtext
\begin{document}
\draft
\title{Anomalous Relaxation 
   in the $XY$ Gauge Glass}

\author{Beom Jun Kim and M.Y. Choi}
\address{
     Department of Physics and Center for Theoretical Physics\\
     Seoul National University\\
     Seoul 151-742, Korea}

\author{S. Ryu and D. Stroud}
\address{Department of Physics\\
          The Ohio State University\\
          Columbus, OH 43210}
\maketitle

\thispagestyle{empty}

\begin{abstract}
To study relaxation dynamics of the two-dimensional $XY$ gauge glass,
we integrate directly the equations of motion and
investigate the energy function.
As usual, it decays exponentially at high temperatures; at low but
non-zero temperatures, it is found to exhibit an
algebraic relaxation.  We compute the relaxation time $\tau$
as a function of the temperature $T$ and find that
the rapid increase of $\tau$ at low temperatures is well
described by $\tau \sim (T-T_g)^{-b}$ with $T_g  = 0.22 \pm 0.02$
and $b =  0.76 \pm 0.05$, which strongly suggests 
a finite-temperature glass transition. 
The decay of vorticity is also examined and 
explained in terms of a simple heuristic model, which attributes
the fast relaxation at high temperatures to annihilation of 
unpinned vortices.
\end{abstract}
\bigskip

\begin{multicols}{2}

Recently, dynamics of the physical systems which have many metastable
states has drawn much interest~\cite{book,mazo,rieger}. In general, such
systems show very slow relaxation at low temperatures, where energy barriers
separating metastable states are sufficiently high compared with the thermal
energy. The $XY$ gauge glass, which has attracted much attention 
in relation to the vortex-glass phase of strongly disordered 
superconductors~\cite{dsfisher91}, is a well-known example of the strongly 
disordered system with a complex energy-landscape.
Such a gauge-glass model is of particular interest in two dimensions, with
regard to the possible glass order at finite temperatures. 
Although it is known that
the two-dimensional (2D) $XY$ model with random bond angles does not display
a finite-temperature Kosterlitz-Thouless (KT) transition in the
strong-disorder regime corresponding to the gauge glass~\cite{gsjeon}, 
there has been  controversy as to the value of the glass transition 
temperature $T_g$: Numerical calculations
of the current-voltage $(IV)$ characteristics~\cite{hyman} have suggested 
a zero-temperature transition. However, the temperature range
probed in Ref.~\onlinecite{hyman} is not sufficient in view of Ref.~\onlinecite{yhli},
where the $IV$ characteristics appear to indicate a 
glass transition at lower temperature $T_g \approx 0.15$.
Further, the analytical result of the vanishing glass-order 
parameter at finite temperatures~\cite{nishimori} does not
exclude the possibility of an algebraic {\em glass} order,
characterized by the algebraic
decay of the {\it glass} correlation function (see Ref.~\onlinecite{huse} 
for the definition). 
This is the counterpart of the usual algebraic order (and the associated 
KT transition) in the 2D pure $XY$ model, where
the Mermin-Wagner theorem~\cite{mermin} does not 
forbid the existence of such a quasi-long range order, i.e., the algebraic
decay of the spin-spin correlation function.
Recently, a similar point has been made on the nature of two-dimensional
disordered lattice~\cite{giamarchi}.
Therefore, we believe that the question on 
the finite-temperature glass transition is not completely settled. Indeed
the possibility of algebraic glass order at low temperatures
has been suggested in Ref.~\onlinecite{sypark}, where it has also been shown
that the numerical calculation of the defect-wall energy~\cite{gingras}
does not rule out the existence of an algebraic glass order at finite
temperatures.

In this work, we study dynamics of the 2D $XY$ gauge glass in the absence
of an external current and investigate
the possible glass order at finite temperatures. For this purpose,
we consider an array of resistively shunted Josephson junctions with random 
magnetic bond angles, whose equations of motion 
are derived from the current conservation condition at each node. 
Via numerical integration, we compute 
the energy function and the vorticity function, and find, as expected,
that both quantities decay exponentially at high temperatures.
In contrast, at low temperatures they are found to exhibit algebraic
relaxation:
Such striking difference in the decay behaviors according to the temperature
$T$ strongly suggests a dynamical glass transition at a finite 
temperature $T_g$. 
The behavior of the energy function allows us to estimate the 
relaxation time $\tau$, giving the temperature dependence 
$\tau \sim (T-T_g)^{-b}$ with $T_g = 0.22 \pm 0.02$ and $b = 0.76 \pm 0.05$. 
Whereas $\tau$ is apparently independent of the system size at sufficiently 
high temperature ($T \gg T_g$), it
increases strikingly with the system size as $T$ approaches $T_g$, 
manifesting its divergent behavior at $T=T_g$ in the thermodynamic limit.
We also present a simple heuristic model to explain the decay of the
vorticity function,
and argue that unpinned vortices play an important role in the relaxation
behavior.

We begin with an $L\times L$ square array of resistively
shunted Josephson junctions 
with periodic boundary conditions in both 
directions. The net current from grain $i$ to grain $j$ is written as the
sum of the Josephson current, the normal current, and the thermal noise current:
\begin{equation} \label{eq:I}
   I_{ij}=I_c\sin(\phi_i-\phi_j-A_{ij})+{V_{ij}\over R}+\Gamma_{ij},
\end{equation}
where $\phi_i$ is the phase of the superconducting order parameter at grain
$i$, $I_c$ is the critical current of the junction,
$V_{ij}$ is the potential difference across the junction, and $R$
is the shunt resistance. The thermal noise current $\Gamma_{ij}$ at temperature 
$T$ is assumed to satisfy 
\begin{equation}
   \langle \Gamma_{ij}(t+\tau)\Gamma_{kl}(t)\rangle=
      {2k_BT\over R}\delta(\tau)(\delta_{ik}\delta_{jl}-
      \delta_{il}\delta_{jk}),
\end{equation}
where $\langle\cdots\rangle$ denotes the ensemble average.
To describe the gauge glass model, the magnetic bond angles $A_{ij}$'s are 
taken  to be quenched random variables distributed uniformly in the 
interval $(-\pi , \pi]$. 
The potential is related to the phase via the Josephson relation,
$d(\phi_i-\phi_j)/dt=2eV_{ij}/\hbar$. This, together with the current 
conservation at each site, allows us to write Eq.~(\ref{eq:I}) in the form 
of a set of $N (\equiv L^2)$ coupled equations:
\begin{equation} \label{eq:motion}
{\sum_j}^\prime \left[ \frac{d}{dt} (\phi_i-\phi_j)+
       \sin(\phi_i-\phi_j-A_{ij})+\gamma\eta_{ij}\right] = 0,
\end{equation}
where the summation is over the nearest neighbors of $i$ and we have introduced
the dimensionless parameter
\begin{equation}
\eta_{ij}\equiv\left({\hbar I_c\over 2ek_BT}\right)^{1/2}{\Gamma_{ij}\over
I_c}\equiv{\Gamma_{ij}\over \gamma I_c}
\end{equation}
and rescaled time $t$ in units of $\hbar/(2eRI_c)$.
It is of interest here to note that Eq.~(\ref{eq:motion}) describes 
the {\em intrinsic}
dynamics of the system, based on the current conservation rule. This is in
contrast to existing studies of dynamical behaviors, where
phenomenological Langevin relaxational 
dynamics or Glauber dynamics has been adopted, starting from the equilibrium 
Hamiltonian~\cite{book,mazo,rieger}.

To study dynamical behaviors, we integrate directly the equations of motion,
given by Eq.~(\ref{eq:motion}) with
$L=8$, 12, 16 and 24. 
In particular, we use random initial conditions with the time step 
$\Delta t = 0.05$, and compute the energy function $E(t)$, the auto-correlation
function $C(t)$, and the vorticity function $v(t)$:
\begin{eqnarray}
& &E(t) \equiv \frac{1}{N}\left\langle\sum_{\langle i,j\rangle}
      \cos[\phi_i(t) - \phi_j(t)-A_{ij}] \right\rangle , \\
& &C(t) \equiv \left\langle \left| \frac{1}{N}\sum_i e^{i[\phi_i(t)-\phi_i(0)]}
\right|^2\right\rangle, \label{eq:Ct} \\
& &v(t) \equiv \frac{1}{N}\left\langle \sum_{{\rm all \;} p } 
   \left|{\sum}^p [\phi_i(t)-\phi_j(t)-A_{ij}]\right|\right\rangle, 
   \label{eq:vt} 
\end{eqnarray}
where $\sum_{\langle i,j\rangle}$ is the summation over nearest-neighboring
pairs, $\langle\cdots\rangle$ denotes the average over both the
initial conditions and disorder configurations. 
The summation ${\sum}^p$ 
is taken in the counter-clockwise direction over the bonds surrounding 
plaquette $p$, and the phase difference $\phi_i -\phi_j -A_{ij}$ is 
defined modulo $2\pi$. 
Since the numerical results are found to be insensitive to
the disorder configuration, one realization of $A_{ij}$ is used 
in most cases. The auto-correlation function in Eq.~(\ref{eq:Ct}) measures
the time correlation of the glass order parameter; 
the vorticity function $v(t)$ in Eq.~(\ref{eq:vt}) measures the average 
chirality on each plaquette regardless of the sign, and is expected 
to be proportional to the average number density of vortices.  
In the limit $t \rightarrow \infty$, we expect 
$E(t)$ and $v(t)$ should approach the equilibrium values 
$E^{eq}$ and $v^{eq}$, respectively, at a given temperature $T$, while $C(t)$
approaches $1/N$ regardless of the temperature (see below). 

\begin{figure}
\centerline{\epsfxsize=7cm \epsfbox{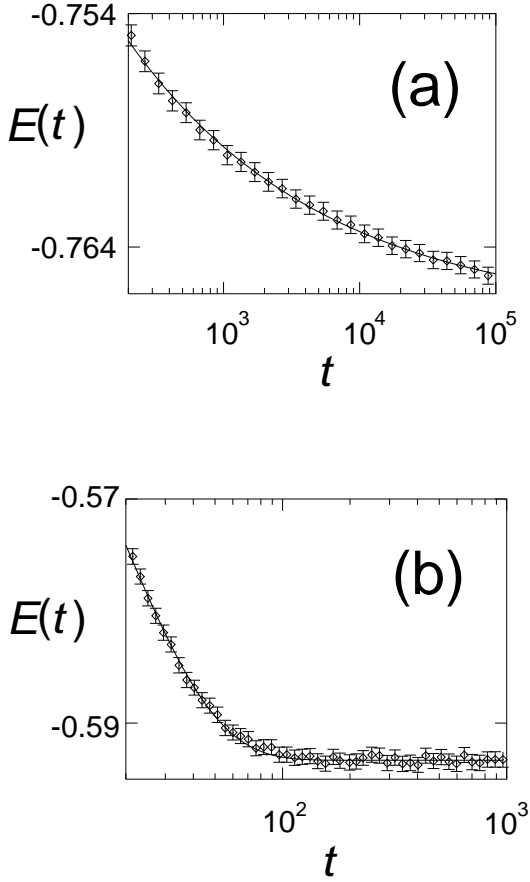}}
\vskip 0.5cm
\caption{ 
Relaxation of the energy function at temperatures (a) $T$=0.1 and
(b) $T$=0.6. The system size is $16\times 16$ and the data have been averaged
over more than $10^3$ samples with different disorder 
configurations and initial conditions, and error bars correspond to two
standard deviations. 
In (a) the solid line represents the algebraic decay
$E(t)-E^{eq} \sim t^{-0.30}$ with $E^{eq}=-0.767$, 
while the solid line in (b) corresponds to $E(t)-E^{eq} \sim \exp(-t/19)$
with $E^{eq}=0.593$. As the temperature is decreased, the decay
behavior changes from the exponential to the algebraic.
}
\label{fig:Et}
\end{figure}

We first summarize the relaxation behavior of the energy function $E(t)$.
Figure~\ref{fig:Et} shows the decay of $E(t)$ (in units of $\hbar I_c/2ek_B$) 
of a $16 \times 16$ array at temperatures (a) $T=0.1$ and (b) $T=0.6$.
At $T=0.1$ it is observed that $E(t)$ does not achieve its equilibrium value
within the time range $t<10^5$, 
which corresponds to $2\times 10^6$ time steps. 
In particular Fig.~\ref{fig:Et}~(a) shows that for $t\gtrsim 200$
$E(t)$ fits well with the algebraic 
decay $E(t) - E^{eq} \sim t^{-0.30}$, 
which is in contrast with the exponential relaxation 
$E(t) - E^{eq} \sim \exp(-t/19)$ in Fig.~\ref{fig:Et}~(b).
This qualitative change in the relaxation behavior may suggest 
a finite-temperature glass transition
with the glass transition temperature $T_g$ lower than 0.6. 
To determine $T_g$ more precisely, we investigate the normalized 
energy function
\begin{equation} \label{eq:tE}
\tilde E(t) \equiv \frac{ E(t) - E^{eq} }{ E(t=0) - E^{eq} },
\end{equation}
defined to satisfy $\tilde E(0) = 1$ and $\tilde E(t\rightarrow \infty) = 0$.
Since a system with a complicated free energy landscape in general 
possesses various scales of the relaxation time~\cite{dotsenko}, 
we write the energy function in the form~\cite{dotsenko,ogielski}
\begin{equation}
\tilde E(t) = \int_0^\infty d\tau^\prime P(\tau^\prime) e^{-t/\tau^\prime},
\end{equation}
where $P(\tau^\prime)$ describes the distribution
of the relaxation time.
This gives the average relaxation time $\tau$ as the integral of
the energy function:
\begin{equation}
\tau \equiv \int_0^\infty d\tau^\prime P(\tau^\prime) \tau^\prime 
  = \int_0^\infty dt' \tilde E(t').
\end{equation}

\begin{figure}
\centerline{\epsfxsize=7cm \epsfbox{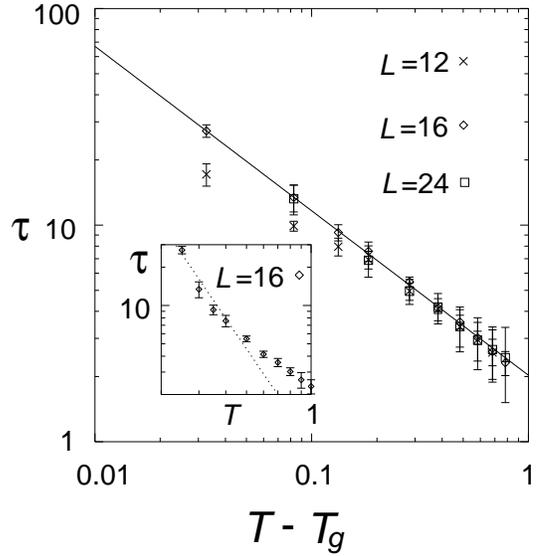}}
\vskip 0.5cm
\caption{
Relaxation time $\tau$ versus the temperature $T-T_g$ in the log-log scale
for system sizes $L=12$, 16, and 24 with $T_g = 0.22 \pm 0.02$. 
Whereas $\tau$ is independent of $L$ at high temperatures,
it strongly depends on $L$ as the temperature is lowered.
The rapid increase of $\tau$ fits well with the form 
$(T-T_g)^{-0.76}$, which is represented by the solid line. 
Inset: $\tau$ versus $T$ in the log-log scale.
The dotted line is the result of the least-square fit 
to the form $T^{-c}$, displaying the incompatibility with the 
zero-temperature glass transition. 
}
\label{fig:tau}
\end{figure}

The obtained behavior of $\tau$ is shown in
Fig.~\ref{fig:tau} for various sizes $L=12$, $16$, and $24$~\cite{error}. 
It is evident that at high
temperatures ($T\gtrsim 0.5$), $\tau$ is independent of $L$ within
error bars.  
At low temperatures, on the other hand, $\tau$ increases with
$L$.  This increase becomes more pronounced as the temperature is lowered,
suggesting divergent behavior of $\tau$ in the thermodynamic limit 
as $T$ approaches $T_g \approx 0.22$ from above.
Such a rapid increase
of the average relaxation time as $T$ is lowered is very well described by the form
\begin{equation}
\tau \sim (T-T_g)^{-b}
\end{equation}
with the glass-transition temperature $T_g = 0.22\pm 0.02$ and the exponent
$b =  0.76\pm 0.05$, which is represented by the 
solid line in Fig.~\ref{fig:tau}. 
The dotted line in Fig.~\ref{fig:tau} is the result of the best fit
to the form $\tau \sim T^{-c}$, i.e., with $T_g = 0$: 
It suggests that the behavior of $\tau$ is incompatible
with the zero-temperature glass transition. 

This method of estimating the transition temperature from the average 
relaxation time has been used in 
the study of the three-dimensional Ising spin-glass model~\cite{ogielski}, 
where both the correlation length in equilibrium
and the average relaxation time in dynamics have been shown to diverge
at the {\em same} transition temperature. 
Indeed a very recent study of the correlation length $\xi(T)$
for the 2D $XY$ gauge-glass model showed that $\xi(T)$ diverges
at $T_g = 0.22\pm 0.01$~\cite{sypark}, 
which is consistent with the value obtained dynamically in this work.  
Here the divergent behavior of the correlation length can be manifested by
characteristic finite-size effects in the numerical study performed
on a system of finite size:
Suppose that the correlation length $\xi$
diverges at $T_g$ like $\xi \sim (T-T_g)^{-\nu}$ with the appropriate
exponent $\nu$.  In the high-temperature regime, the system size $L$
is in general much larger than $\xi$, and
the system is expected not to display appreciable size dependence.
Near $T_g$, on the other hand, we have
$\xi \gtrsim L$, which leads to strong finite-size effects. 

\begin{figure}
\centerline{\epsfxsize=7cm \epsfbox{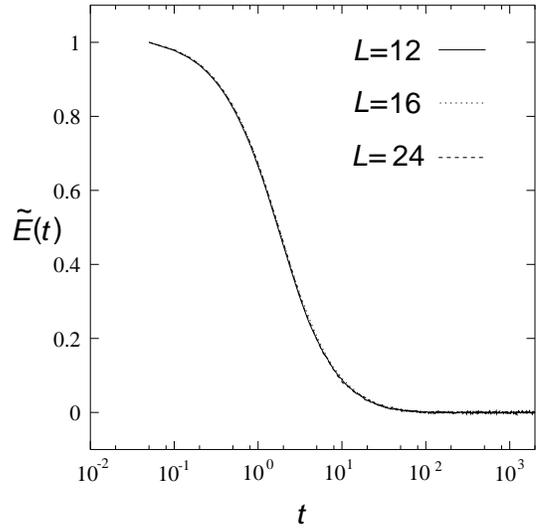}}
\vskip 0.5cm
\caption{ 
Relaxation of the normalized energy function $\tilde E$ at temperature 
$T=0.6$.  It is evident that $\tilde E(t)$ is independent of the system size.
Typical errors are quite small, of the order of $10^{-3}$.
}
\label{fig:tE0.6}
\end{figure}

\begin{figure}
\centerline{\epsfxsize=5cm \epsfbox{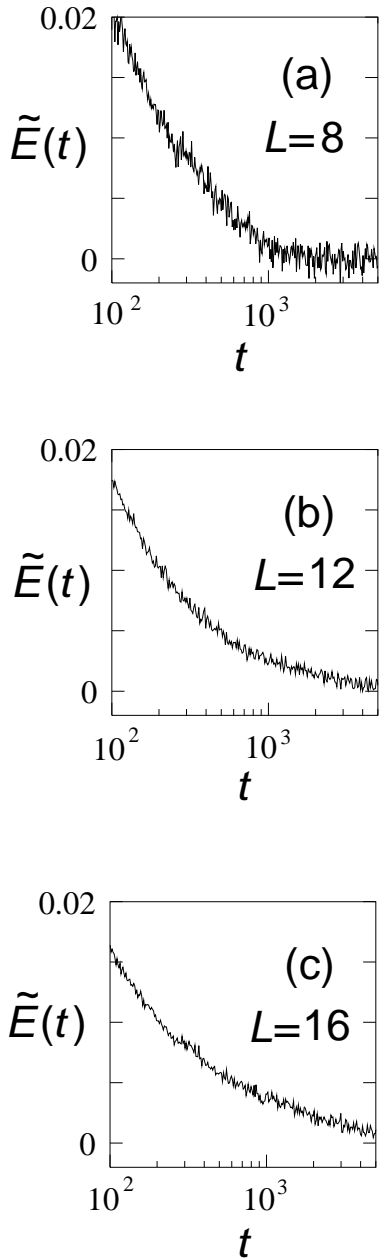}}
\caption{ 
Relaxation of the normalized energy function $\tilde E$ at temperature 
$T=0.25$ for sizes (a) $L=8$, (b) $L=12$, and (c) $L=16$. 
As the system size is increased, the relaxation becomes slower.
}
\label{fig:tE0.1}
\end{figure}

To examine such behavior, we compute
the normalized energy function $\tilde E(t)$ at $T=0.6$ and at $T=0.25$ 
and show the results in Fig.~\ref{fig:tE0.6} and in Fig.~\ref{fig:tE0.1},
respectively.  
At high temperatures ($T=0.6$) Fig.~\ref{fig:tE0.6} shows that 
the decay of $\tilde E(t)$ does not depend on the system size. 
In contrast, at low temperatures ($T=0.25$)
the relaxation becomes slower as the size is increased,
as displayed in Fig.~\ref{fig:tE0.1}.  
This size-dependence becomes more distinct as the temperature approaches
$T_g \approx 0.22$~\cite{comm}.
These different size-dependent behaviors of $\tilde E(t)$ 
according to the temperature are certainly 
consistent with the divergence of the
correlation length at $T_g \approx 0.22$, which indeed supports that
both the correlation length in equilibrium
and the average relaxation time in dynamics diverge
at the {\em same} transition temperature $T_g \approx 0.22$. 
This estimation is somewhat higher than the value $T_g \approx 0.15$ 
estimated from the $IV$ characteristics~\cite{yhli}.
The very slow relaxation at low temperatures makes it
difficult to reach the stationary state
in numerical simulations, and presumably causes the 
discrepancy in determining $T_g$. 

\begin{figure}
\centerline{\epsfxsize=7cm \epsfbox{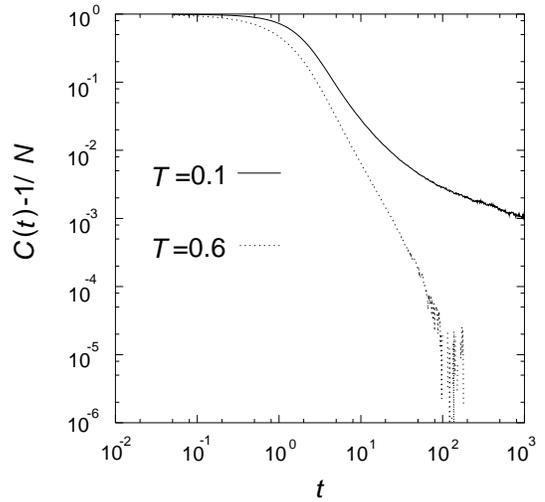}}
\vskip 0.5cm
\caption{ Relaxation of the auto-correlation function $C(t)$ at temperatures
$T=0.1$ and 0.6 for $16\times 16$ array ($N=256$). Since $C(t)\rightarrow 1/N$
as $t\rightarrow \infty$, $C(t) - 1/N$ is plotted as a function of
time $t$.
}
\label{fig:Ct}
\end{figure}

We next investigate the auto-correlation function defined in 
Eq.~(\ref{eq:Ct}), 
which can be cast in the form:
\end{multicols}
\noindent\rule{0.5\textwidth}{0.1ex}\rule{0.1ex}{2ex}\hfill
\widetext
\begin{eqnarray}
C(t) &=& \left\langle \frac{1}{N^2}\sum_{i,j}\cos[\phi_{ij}(t)-\phi_{ij}(0)]
\right\rangle \nonumber \\
& =& \frac{1}{N}+\frac{1}{N^2}\left\langle \sum_{i\neq j}[\cos\phi_{ij}(t)
\cos\phi_{ij}(0) + \sin\phi_{ij}(t)\sin\phi_{ij}(0)]\right\rangle
\end{eqnarray}
\hfill\raisebox{-1.9ex}{\rule{0.1ex}{2ex}}\rule{0.5\textwidth}{0.1ex}
\begin{multicols}{2}
\narrowtext
\noindent
with $\phi_{ij}(t) \equiv \phi_i(t)-\phi_j(t)$. At $t=0$, we have the
auto-correlation function $C(t=0) = \frac{1}{N} + \frac{1}{N^2}N(N-1) = 1$. For
$t\rightarrow \infty$, we expect that $\phi_{ij}(t)$ and $\phi_{ij}(0)$ become
independent of each other:
\begin{equation}
\langle \cos\phi_{ij}(t)\cos\phi_{ij}(0)\rangle \rightarrow
\langle \cos\phi_{ij}(t)\rangle \langle \cos\phi_{ij}(0)\rangle , 
\end{equation}
which, combined with $\langle \cos\phi_{ij}(0)\rangle = 0$ upon averaging over 
initial configurations, reveals that $C(t\rightarrow \infty) = 1/N$ at any
temperature. The auto-correlation function obtained numerically at
temperatures $T=0.1$ and $0.6$ is plotted in Fig.~\ref{fig:Ct}, which 
shows that approach of $C(t)$ to the asymptotic value for $T < T_g$ is 
qualitatively different from that for $T > T_g$. 

We finally study the decay of the vorticity function $v(t)$,
and propose a heuristic model to understand the nature of the
transition in terms of annihilation 
of vortices and antivortices. The vortices present in the system may be 
classified into three types with regard to their origins: 
quenched vortices induced by the random bond angle (type I), 
thermally generated vortices (type II), and vortices introduced by the
random initial conditions (type III). 
It is plausible to assume that the numbers of vortices and antivortices are
equal for each type.
Since vortices of type III annihilate as time goes on, 
only vortices of types I and II exist in equilibrium; 
thermal vortices (of type II) exist at nonzero temperatures 
($T>0$)~\cite{gsjeon}.
It is further assumed that vortices of type III annihilate with 
antivortices 
of the same type or {\em unpinned} antivortices of types I and II. 
Since a vortex and an antivortex attract each other, 
and annihilate when they meet, we suppose that a vortex-antivortex pair 
with the size (i.e., the distance between the vortex and the antivortex) 
smaller than $l$ annihilates after time ${\bar t}$.
To obtain the transient behavior of the vorticity function, 
we consider a vortex configuration consisting of $n$ antivortices of 
type III and $n_0$ unpinned antivortices of types I and II. 
The average distance $a$ between the randomly positioned antivortices is 
given by $a \approx \sqrt{S/\pi (n + n_0)}$,
where $S$ is the total area. 
We now consider a single vortex of type III added at a random position.
The probability $P$ of the distance between this additional vortex and the 
nearest antivortex being smaller than $l$ may be written as
$P \approx \pi l^2/a^2 \approx (\pi^2 l^2/S)(n+n_0)$, 
which leads to the decay rate for $n$ vortices of type III
\begin{equation} \label{eq:dndt}
 \frac{dn}{dt} \approx -\frac{\pi^2 l^2}{t_0 S}\, n(n+n_0).
\end{equation}
Note here that $n_0$ has been assumed to take its equilibrium value and 
the fluctuations of $n_0$ have been neglected. 
Thus Eq.~(\ref{eq:dndt}) is valid for $t$ greater than the ``equilibration'' 
time $t_0$ of $n_0$ beyond which $n_0(t)$ reaches 
its equilibrium value $n_0$.
It is straightforward to obtain the solution of Eq.~(\ref{eq:dndt}):
$n(t)=(n_0/2)$ $[\coth(c_1t{+}c_2)$ -$1]$,
where $c_1 \equiv n_0\pi^2 l^2/2{\bar t} S$ and 
$c_2 \equiv (1/2) \log [1+n_0 /n({\bar t})] - c_1 {\bar t}$. 
The decrease of the vortices of type III results in the relaxation of
the vorticity function.  Therefore $n$ is proportional to $v-v^{eq}$, 
and the vorticity function is expected to exhibit the behavior
\begin{equation} \label{eq:coth}
 v(t) - v^{eq} \sim  [\coth(c_1t+c_2) - 1],
\end{equation} 
again for $t$ is larger than $t_0$.

\begin{figure}
\centerline{\epsfxsize=7cm \epsfbox{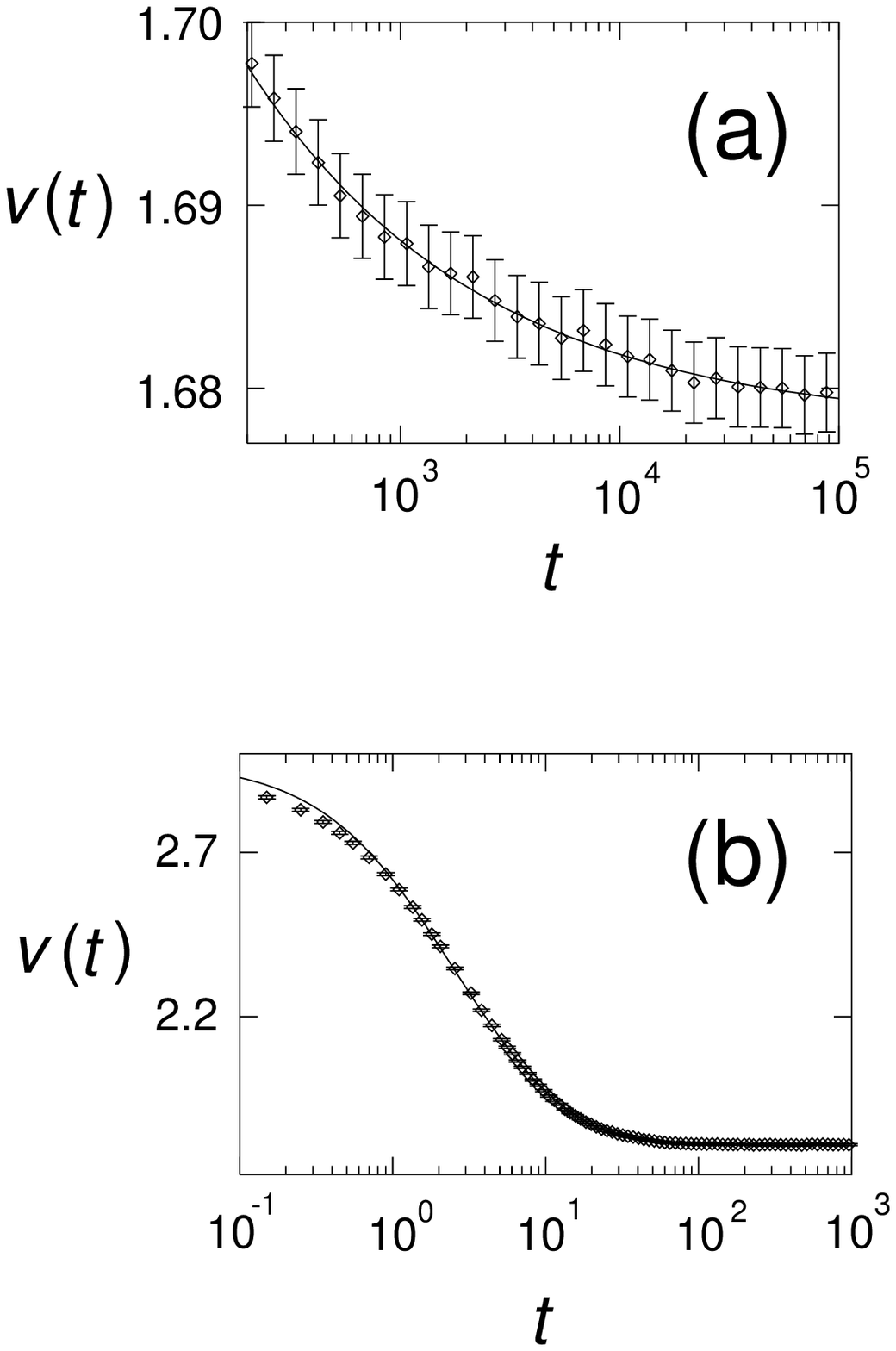}}
\vskip 0.5cm
\caption{ 
Relaxation of the vorticity function at temperatures (a) $T$=0.1 and
(b) $T$=0.6.  The system size and other conditions are the same as those in 
Fig.~\ref{fig:Et}.
In (a) the solid line represents $v(t) - 1.678 \sim t^{-0.41}$, while the solid line in (b)
corresponds to $v(t) - 1.702 \sim \coth(0.035t+0.087)-1$ [see Eq.~(\ref{eq:coth})].  
Note the difference in the 
vertical scale.
}
\label{fig:vt}
\end{figure}

We compute numerically the vorticity function, and exhibit the results in
Fig.~\ref{fig:vt} for (a) $T=0.1$ and (b) $T=0.6$. 
At high temperatures ($T=0.6$), Fig.~\ref{fig:vt}~(b) shows that 
the numerical data indeed fit very well with Eq.~(\ref{eq:coth}) 
with $c_1 \approx 0.035$ and $c_2 \approx 0.087$, even for rather small 
$t$ ($\gtrsim 1$).
On the other hand, at low temperatures, the equilibration
time $t_0$ grows large and is expected to be larger than the 
observation time $t$, thus making
Eqs.~(\ref{eq:dndt}) and (\ref{eq:coth}) invalid. 
In this case the decay of the vorticity function is described
by an algebraic function rather than Eq.~(\ref{eq:coth}).
In Fig.~\ref{fig:vt} (a), the vorticity function at $T=0.1$
is shown to be well described by the algebraic function 
$v(t) - v^{eq} \sim t^{-0.41}$.
We have also measured the average relaxation time 
$\tau_{\rm vor}$ of the normalized vorticity function,
defined similarly to Eq.~(\ref{eq:tE}), and found that 
it is also divergent near $T_g$, again supporting 
a finite-temperature glass transition.  

In summary, we have investigated dynamics of a two-dimensional $XY$ 
gauge glass via numerical integration of equations of motion 
obtained from current conservation conditions.
At low temperatures both the energy function and the vorticity function
have been found to relax algebraically, which is in contrast with 
the exponential decay at high temperatures. 
The auto-correlation function at low temperature has also been shown to
relax very slowly. 
The relaxation time $\tau$ has been computed and
shown to follow $\tau \sim (T-T_g)^{-0.76}$ with $T_g = 0.22\pm 0.02$. 
Such divergence of the relaxation time in turn implies the divergence
of the correlation length, or slowly (e.g., algebraically)
decaying spatial correlations~\cite{sypark}.
Finally, the relaxation of the vorticity function has been shown
to be closely related to diffusion of unpinned vortices.

BJK and MYC thank S.J. Lee for useful discussions, and acknowledge the
partial support from the BSRI Program, Ministry
of Education of Korea, and from the KOSEF through the SRC program.  
SR and DS was supported in part by the NSF Grant No. DMR94-02131 
and by the DOE Grant DE-FG02-90ER-45427.

\vskip 1cm

\vskip 1cm

\end{multicols}

\begin{thebibliography}{99}
\bibitem{error} Numerically, we compute 
 $\tau(t) \equiv \int_0^t dt' \tilde E(t')$ for sufficiently large times
 at which $E(t)$ reaches the equilibrium value 
 (i.e., $\tilde E(t) \approx 0$), and take the time-average of $\tau(t)$
 to obtain the average relaxation time.  The error bars
 are then estimated from the fluctuations of $\tau(t)$ in this large-time
 regime.
\bibitem{comm} These finite-size effects eventually disappear 
as the size is increased further (so that $L$ is sufficiently 
larger than $\xi(T)$ at the given temperature). 
For example, we have checked the behavior of $\tau$ at $T=0.3$, 
and found that it increases 
with the size up to $L=16$ and then saturates (see Fig.~\ref{fig:tau}).
\end{thebibliography}
\end{document}